\documentclass[aps,prx,twocolumn,superscriptaddress]{revtex4-2}

\usepackage{amsmath}
\usepackage{amssymb}
\usepackage{amsthm}
\usepackage{mathtools}

\usepackage{xcolor}
\usepackage{import}

\usepackage{braket}

\usepackage[normalem]{ulem}

\usepackage[colorlinks,bookmarks=false,citecolor=blue,linkcolor=red,urlcolor=blue]{hyperref}

\begin{document}


\title{In-Gap Band Formation in a Periodically Driven Charge Density Wave  Insulator}
\author{Alexander Osterkorn}
\thanks{These authors contributed equally.}
\email{osterkorn@theorie.physik.uni-goettingen.de}
\affiliation{Institute for Theoretical Physics, Georg-August-University G\"ottingen, Friedrich-Hund-Platz 1, D-37077 G\"ottingen, Germany}
\author{Constantin Meyer}
\thanks{These authors contributed equally.}
\affiliation{Institute for Theoretical Physics, Georg-August-University G\"ottingen, Friedrich-Hund-Platz 1, D-37077 G\"ottingen, Germany}
\author{Salvatore R. Manmana}
\email[]{salvatore.manmana@uni-goettingen.de}
\affiliation{Institute for Theoretical Physics, Georg-August-University G\"ottingen, Friedrich-Hund-Platz 1, D-37077 G\"ottingen, Germany}






\date{\today}

\begin{abstract}
Periodically driven quantum many-body systems host unconventional behavior not realized at equilibrium. 
Here we investigate such a setup for strongly interacting spinless fermions on a chain, which at zero temperature and strong interactions form a charge density wave insulator. 
Using unbiased numerical matrix product state methods for time-dependent spectral functions, 
we find that driving of the correlated charge-density wave insulator leads not only to a renormalization of the excitation spectrum as predicted by an effective Floquet Hamiltonian, but also to a cosine-like in-gap feature. 
This is not obtained for a charge density wave model without interactions.
A mean-field treatment provides a partial explanation in terms of doublon excitations. 
However, the full picture needs to take into account strong correlation effects.  
\end{abstract}

\maketitle

\section{Introduction\label{sec:intro}}

A central driving force of modern condensed matter physics is the realization of novel phases of matter out-of equilibrium like transient superconductivity~\cite{mitrano_2016_cdw_neq_sc,fausti_2011_light_induced_sc,hu_2014_neq_sc,wang_2018_lightinduced_superconductivity,Paeckel2020} or excitonic insulators in transition metal dichalcogenides~\cite{Erben2018,jin_2018_tmdcs_heterostructures_excitons,zhu_2017_tmdcs_excitons,kunstmann_2018_interlayer_excitons,jin_2019_moire_excitons,chernikov_2015_excitons,hellmann_2012_cdw}.
These are typically created from a highly complex interplay of band structure, (electronic) interactions, and excitations by a light-field.
Nowadays, experimental techniques allow to actively ``engineer'' properties of  many-body quantum systems in such out-of-equilibrium systems in a highly controlled way \cite{de2021colloquium,oka2019floquet,rubio_kennes_arxiv}.
This opens up the possibility to create behavior that is not even possible in equilibrium setups.
An important pathway is to induce excitations whose interplay with the electronic interactions can lead to intriguing transient behavior.
Such excitations can be realized in experiments, e.g., by ultrashort laser pulses in so-called pump-probe setups~\cite{review_krausz,Freericks2009,ejima2022photoinduced}, or by continuous periodic driving of the systems, e.g., by shaking ultracold atom systems on optical lattices~\cite{review_bloch,Bloch2005,bloch_2012_ultracold_quantum_gases_review,aidelsburger_2011_optical_superlattices_magnetic_field,aidelsburger_2013_optical_superlattices_magnetic_field,schweizer_2016_superlattice_spin_pumping_spin_current,trotzky_2008_optical_lattice_superexchange}.
A much studied theoretical idealization is (infinite) periodic driving, which can be addressed by Floquet theory.
In this framework the properties of the system can be described by a time-independent effective Hamiltonian~\cite{Bukov2015,Eckardt2017}.
Control over the parameters of the driving translates to control over the effective Hamiltonian and this, in turn, allows to manipulate order parameters~\cite{Kennes2018}, induce topological order~\cite{KitagawaDemler} etc.

It is possible to derive 
time-independent approximate effective Hamiltonians in the low-~\cite{Vogl2020b} and high-frequency~\cite{Bukov2015,Eckardt2015} regimes, and development of methods in this vein is ongoing~\cite{Vogl2019}. 
In addition, the periodic driving generically leads to energy absorption evolving the system towards an infinite-temperature state~\cite{lazarides_PRE2014,dalessio_PRX2014}.
This restricts the relevance of such effective Hamiltonians to regimes in which the energy absorption is suppressed like the high-frequency regime.
Additional transient dynamics can be induced by the switching-on procedure~\cite{kuwahara2016floquet,Eckardt2017,Kalthoff2018,novivcenko2017floquet,novivcenko2022flow}.
One major topic of interest is the role of interactions in strongly driven systems.
It has been shown that in Hubbard systems at resonance the interaction can be renormalized, and double occupancies can be enabled~\cite{Eckardt2015,Bukov2015,Herrmann2017}; one can even tune the parameters of the driving so that the fermions behave like free particles~\cite{Bukov2016}.  


In our work, we 
investigate driven strongly interacting fermions including a sudden switching-on procedure and without assuming the high-frequency approximation.
This allows us to look for new effects beyond these limiting cases.
Therefore, we calculate non-equilibrium spectral functions with unbiased matrix product state (MPS)~\cite{Schollwock2011,Paeckel2019} approaches,
and systematically investigate correlation effects by comparing to non-interacting and mean-field scenarios.
This is in contrast to other approaches, which e.g. rely on the Floquet-Magnus expansion~\cite{Bukov2015,Bukov2016} in terms of the inverse driving frequency.
We focus on a simple paradigmatic model for strongly correlated physics, namely a chain of spinless fermions with nearest-neighbor interactions.
At half filling and zero temperature the model is known to undergo a Berezhinskii-Kosterlitz-Thouless (BKT) type transition~\cite{sachdev1999quantum} from a Luttinger liquid (LL)~\cite{giamarchi} to a interacting charge density wave (CDW) insulator~\cite{book_gebhard} when increasing the interaction strength.
Driving the system with frequencies much larger than the gap (``Magnus case''), a renormalization of the gap size for this system is predicted~\cite{Kennes2018}.
We investigate this by calculating  non-equilibrium single-particle spectral functions.
For a static band structure upon monochromatic periodic driving, one expects on general grounds the formation of Floquet side-bands~\cite{Wang2013}. 
Our approach allows us to study both effects, but also to go beyond and to look for the emergence of additional spectral features.

\section{Driving a Strongly Correlated Charge-Density-Wave Insulator\label{sec:model}}

\begin{figure*}
\includegraphics[width=\textwidth]{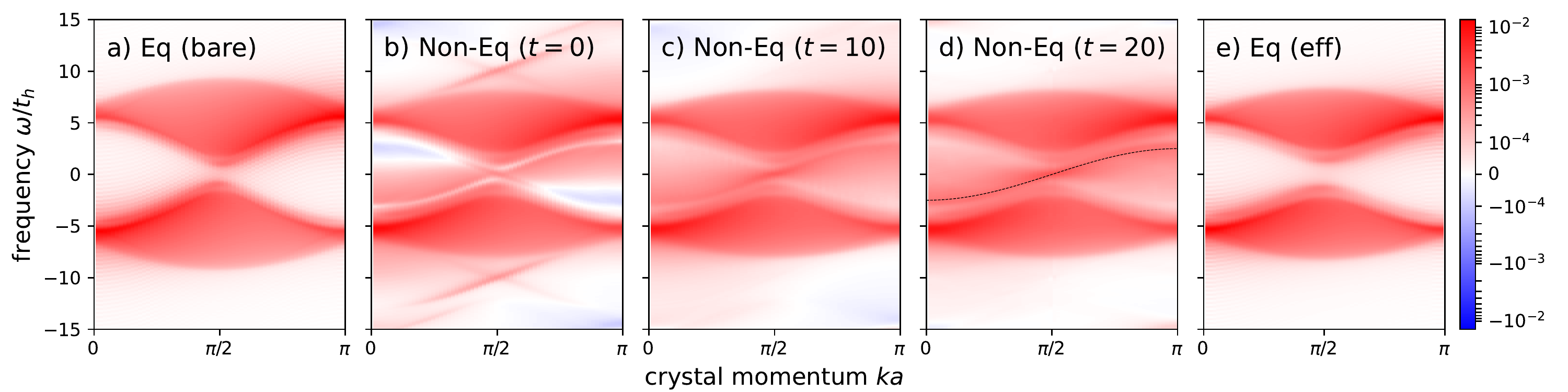}
\caption{(Non-)Equilibrium spectral functions  of the periodically driven CDW ground state at $V/t_h = 5$.
a) and e) show the equilibrium spectral functions $A_k(\omega)$ of the system with bare and renormalized parameters, respectively.
b)-d) show the non-equilibrium spectral functions $A_k(t, \omega)$ upon driving at the instances indicated.
Times are measured in units $t_h^{-1}$.
The equilibrium spectral features in e) are also present in the spectral function of the driven system.
In addition, there is clearly additional spectral weight appearing whose main feature is well approximated by $f(k) \approx -2.5 \cos(k a)$ (dashed line in d)).
All data is obtained with MPS time-evolution for a system with $L = 64$ chain sites and open boundary conditions.}
\label{fig:specfunc_heatmaps_comp_eff}
\end{figure*}


We consider a periodically driven 
chain of interacting spinless fermions described by the Hamiltonian
\begin{align}\begin{split}
 \hat H(t) &= -t_h \sum_{i = 1}^{L-1} \left( e^{i A(t)} c_i^\dagger c^{\phantom{\dagger}}_{i+1} + \text{H.c.} \right) \\
  &\quad + V \sum_{i=1}^L \left( n_i - \frac{1}{2} \right) \left( n_j - \frac{1}{2} \right) \\
  &\quad + \sum_{i=1}^L \mu_i n_i 
  \, ,
\end{split}\end{align}
where $V$ is the strength of the density-density interaction and $\mu_i$ is an on-site potential. $A(t) = \theta(t) A_0 \sin\big( \Omega t \big)$ is a time-dependent vector potential, which is switched on at time $t=0$.
This type of coupling is known as \emph{Peierls substitution}~\cite{Peierls1933} and models a monochromatic classical light-field, which couples to the electrons in the system (here: spinless fermions).
We will mostly consider open boundary conditions (OBC) and for comparison periodic boundary conditions (PBC).
For $A(t=0)$ Bethe ansatz (BA)\cite{DesCloizeaux1966} gives the BKT transition at $V/t_h = 2$.
In the following, we will consider driving a system in the CDW phase at $V/t_h = 5$, for which the energy gap according to BA~\cite{DesCloizeaux1966} is $\Delta/t_h \approx 1.576$.

We define the non-equilibrium generalization of the spectral function via the Fourier transform of the retarded Green's function 
\begin{align}\begin{split}
 A^{\text{ret}}_k(t, \omega) &= -\operatorname{Im} \frac{1}{\sqrt{2\pi}} \int_{-\infty}^\infty \text{d}\tau \; e^{(i\omega - \eta) \tau} G_{kk}^{\text{ret}}(t, \tau), \\
 G_{\alpha \beta}^\text{ret}(t', \tau) &:= -i \theta\left( \tau \right) \big\langle \big\{ c_\alpha(t' + \tau), c_\beta^\dagger(t') \big\} \big\rangle \, 
\end{split}\end{align}
with a damping factor $\eta \approx 0.1$ as further explained in App.~\ref{app:details_gf}.
Integration of $A_k ^{\rm ret}(t,\omega)$ over the crystal momenta $k$ directly yields the time-dependent density of states (tDOS). 
This quantity relates to measurements in time-dependent angle resolved photoemission spectroscopy (trARPES), although a more detailed modelling is required for a direct comparison to experiments~\cite{Freericks2009,Freericks2015,Freericks2016}, as well as for its interpretation in the deep nonequilibrium regime \cite{Kalthoff2019}. 
Nevertheless, these quantities allow us to study qualitative changes of the spectral function with time, such as the formation of additional branches, or a change of the band structure due to the excitation.
 

We complement our study by considering the time evolution of the CDW order parameter
\begin{equation}
 \mathcal{O}_\text{CDW}(t) := \frac 2 L \sum_{i~ \text{even}} \left(n_{i+1}(t) - n_i(t) \right) \, ;
 \label{eq:orderparameter}
\end{equation}
further details are explained in  App.~\ref{app:details_gf}.  
We keep also track of the time evolution of the energy, $E(t) = \langle H(t) \rangle$, which serves as a measure for the heating of the system.  



\section{Results\label{sec:results}}

\begin{figure}
\includegraphics[width=0.5\textwidth]{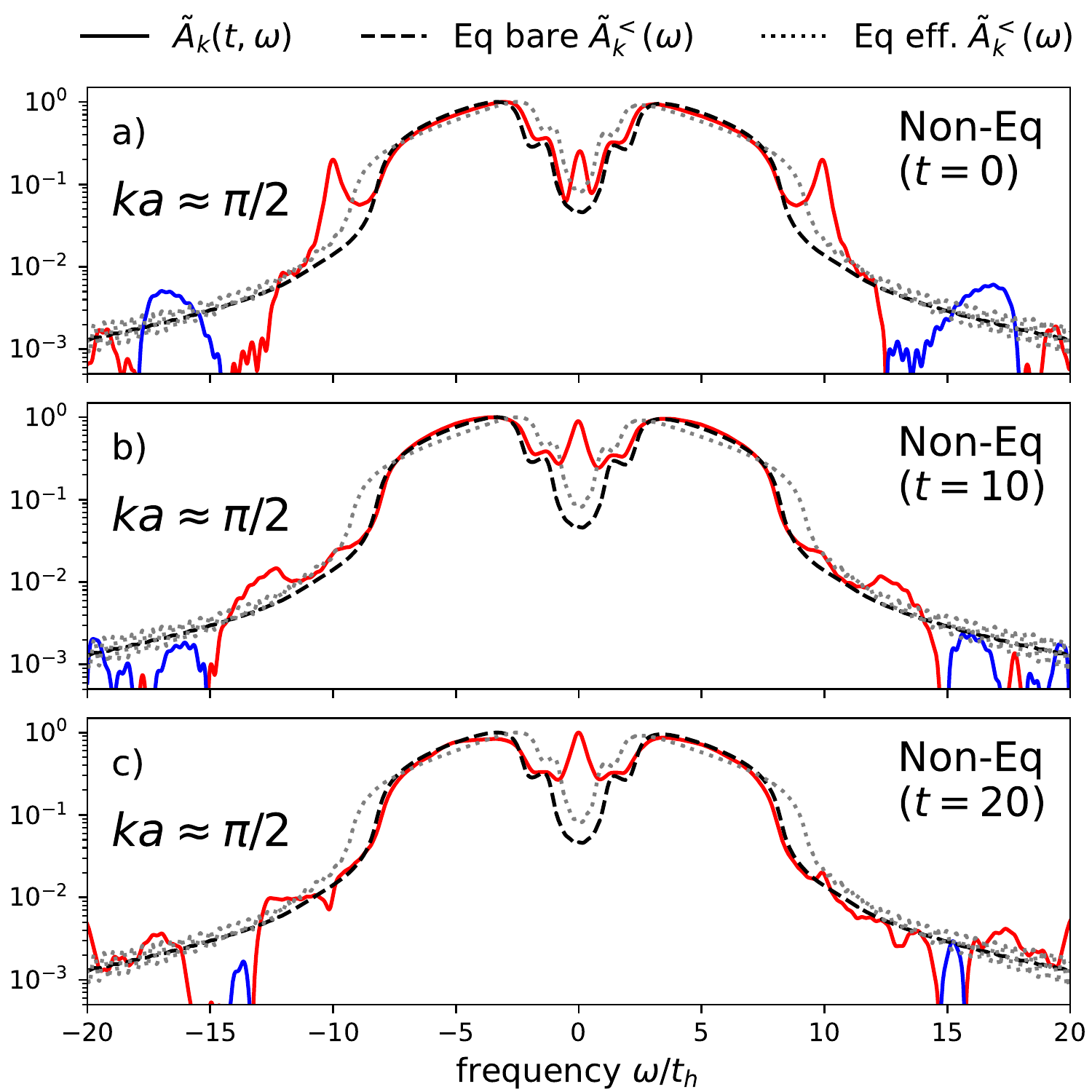}
\caption{Cross section plots at $k \approx \frac{\pi}{2}$ through the data in Fig. \ref{fig:specfunc_heatmaps_comp_eff}.
The dotted grey lines present the equilibrium spectral function of the undriven system, the dashed black lines correspond to the system with Floquet-renormalized parameters.
The data shown is normalized to the maximum value at this $k$-slice: $ \tilde{A}_k(t,\omega) = A_k(t,\omega) \; / \; \max_{\omega} \, A_k(t,\omega)$.
The solid lines correspond to the non-equilibrium spectral functions (red color = positive, blue color = negative).
\label{fig:cross_sec}}
\end{figure}

For periodic driving well above the band gap (``Magnus case'') the properties of the driven model are expected to be well-described by an effective Hamiltonian according to the Floquet-Magnus expansion \cite{Eckardt2015,Kennes2018}.
It is given by the original Hamiltonian but with a renormalized hopping parameter $t_h^\text{eff} = J_0(A_0) t_h \approx 0.7652 t_h$ for $A_0 = 1$ as in Ref.~\onlinecite{Kennes2018}.
This corresponds to a model with $V / t_h^\text{eff} \simeq 1.3V/t_h$.
In the following, we consider these aspects by studying the system at $V/t_h = 5$, which is deep in the CDW insulating phase.


 \begin{figure*}
\includegraphics[width=\textwidth]{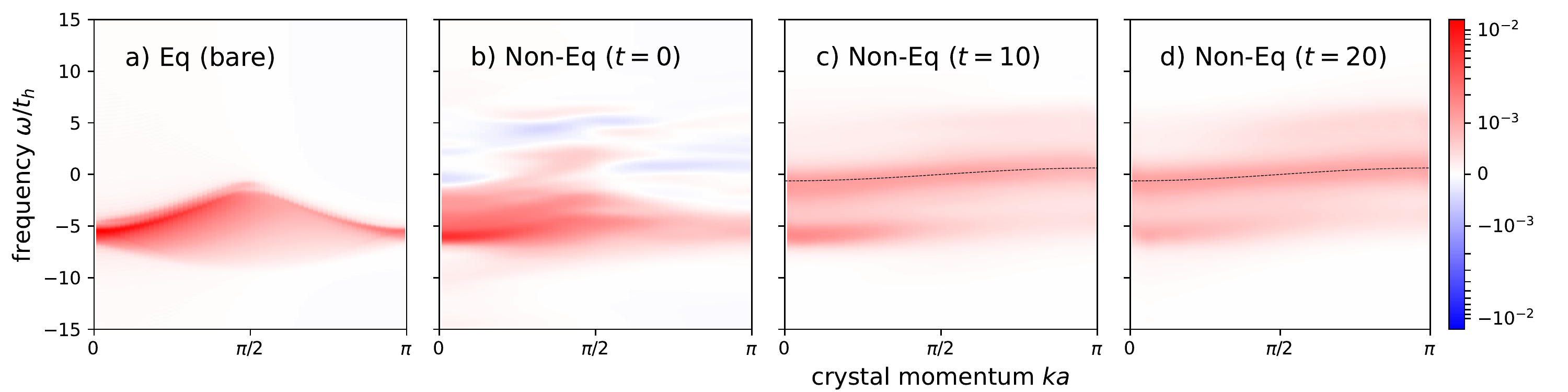}
\caption{Lesser contribution $A_k^{\text{ret}, <}(t, \omega)$ to the non-equilibrium spectral function for $\Omega \approx 4.2 t_h$ and $V/t_h = 5$.
The data in this figure has been obtained for a chain length of $L = 64$ sites and for OBC.
Here $f(k) \sim -\frac{2.5}{4}\cos(ka)$ (dotted line) roughly describes the central feature.
\label{fig:specfunc_heatmap_obc64_om42}}
\end{figure*}

Figs.~\ref{fig:specfunc_heatmaps_comp_eff} and~\ref{fig:cross_sec} show results for the equilibrium and for the non-equilibrium spectral functions at waiting times $t=0, \, 10 t_h^{-1},$ and $20 t_h^{-1}$.
The driving frequency of $\Omega/t_h = 10 > \Delta/t_h \approx 1.576$ is larger than the spectral gap but not substantially larger than the interaction strength so that we are not deeply in the Magnus regime.
Fig.~\ref{fig:specfunc_heatmaps_comp_eff}(e) shows the equilibrium spectral function for the effective Hamiltonian with the renormalized hopping matrix element.
Let us consider the equilibrium results first:
In Fig.~\ref{fig:specfunc_heatmaps_comp_eff}(a) we can identify the equilibrium continuum of excitations~\cite{Pereira2009} as well as the spectral gap located around $\omega = 0$.
Despite finite size effects its minimal size around $k = \pi/2$ is in agreement with the BA prediction.
The spectral function of the effective Hamiltonian in Fig.~\ref{fig:specfunc_heatmaps_comp_eff}(e) looks similar to Fig.~\ref{fig:specfunc_heatmaps_comp_eff}(a) but the width of the continuum is smaller and the gap is larger.
Turning to the non-equilibrium results,
we see that the spectral function at waiting time $t = 0$ in Fig.~\ref{fig:specfunc_heatmaps_comp_eff}(b) looks quite similar to  the equilibrium result for the effective Hamiltonian in Fig.~\ref{fig:specfunc_heatmaps_comp_eff}(e), but it possesses additional features.
In particular, a new in-gap band comes into appearance and the continuum changes slightly its size and form.
Around frequencies $\omega/t_h = \pm 10$ weak signals appear, which seem to echo the in-gap feature.
These are reminiscent of Floquet sidebands, which are observed in time-resolved ARPES experiments~\cite{Wang2013}. 
It is noteworthy, however, that one seems to obtain these ``echoes'' only for the in-gap signal, but not for the main spectral features.
At later waiting times $t = 10 t_h^{-1}$ and $t = 20 t_h^{-1}$, the shape of the continuum does not further change, but the in-gap signal becomes more pronounced.
This is further confirmed by looking at Fig.~\ref{fig:cross_sec}, which shows a momentum cut through the central region of the spectral function at $k \approx \pi/2$ for waiting times $t = 0$, $10 t_h^{-1}$ and $20 t_h^{-1}$.
In order to better focus on the relative distribution of spectral weight we normalized all spectral functions to their maximum value at that $k$-slice.
In all cases, we identify two main lobes and smaller peaks.
Let us follow the behavior of the main lobes and of the largest peaks at $\omega = \pm 10$ and $\omega = 0$:
At waiting time $t = 0$, the lobes show a small difference between the nonequilibrium result and the result of the effective Hamiltonian, which becomes even smaller at later waiting times.
It is noteworthy that the similarity to the effective description is already obtained at waiting time $t = 0$, although the effective Hamiltonian relied on the infinite-driving assumption.
The additional signals at $\omega/t_h = \pm 10$ oscillate in time, but are suppressed with increasing time.
Nevertheless, on the time scale treated by us, the peak at $\omega = 0$ becomes more pronounced with time and so the cosine-like in-gap feature of Fig.~\ref{fig:specfunc_heatmaps_comp_eff} appears stable on the time scales treated by us.
We checked that it is also present for other system sizes and periodic boundary conditions (PBC) so that boundary effects can be ruled out as an explanation (cf. App.~\ref{app:supp_figures}).
By comparing results for $L=32$ and $L=64$ we find that the peak gets sharper for larger system size, while keeping the relative weight. 
At early waiting times negative weight appears in the spectral function.
This is not an artefact and traces back to the non-equilibrium nature of the state.
It was reported recently \cite{Kalthoff2018, Uhrig2019} that upon averaging of the Wigner average time coordinate over a driving period, the non-equilibrium density of states for fermions can be shown to be positive.
However, at later waiting times away from the turning-on of the field at time $t = 0$, our spectral function (obtained using ``horizontal time coordinates'' \cite{Kalthoff2018}) is also almost completely positive without this procedure.


  
\begin{figure}[b]
\includegraphics[width=0.5\textwidth]{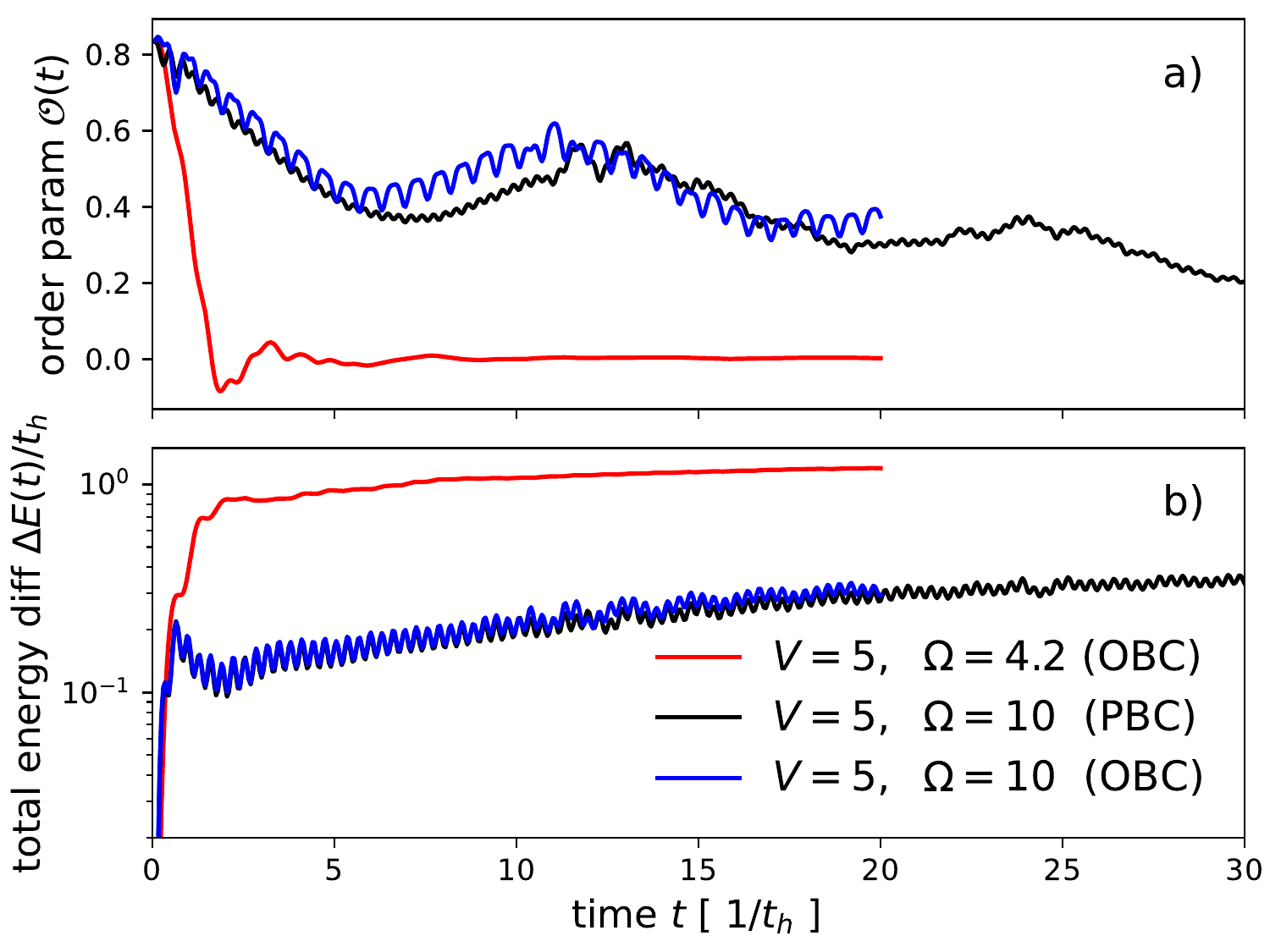}
\caption{Dynamics of the total energy and the order parameter after the sudden switch-on of the driving field.
In order to be able to see the order parameter dynamics, this data is obtained for a model with a pinning potential (as outlined in App.~\ref{app:details_gf}) that leads to an explicit breaking of the translational symmetry.
The chain length is $L = 32$ and both periodic and open boundary conditions are presented.
\label{fig:tevol_en_orderparam}}
\end{figure}

To better understand our findings, we study in Fig.~\ref{fig:specfunc_heatmap_obc64_om42} the system at the same value of $V/t_h = 5$ but with a driving frequency closer to resonance $\Omega/t_h \approx 42 \cdot 10^{-1}$ 
for comparison. 
The additional feature in the gap region in this case is even stronger pronounced and it goes hand in hand with a significant reduction of the original spectral features of the CDW insulator.
The question arises how this disappearance of the quasiparticle continuum is connected with a destruction of the CDW state.
To study this, we calculate the time-evolution of the CDW order parameter $\mathcal{O}_{\textrm{CDW}}(t)$, which is displayed in Fig.~\ref{fig:tevol_en_orderparam} for driving in the Magnus regime and closer to resonance.
In the latter case, $\mathcal{O}_{\textrm{CDW}}(t)$ completely vanishes on a time scale $t \approx 10 t_\text{h}^{-1}$, which is in agreement with the time scale on which the holon continuum disappears in the spectral function.
The behavior for $\Omega/t_h = 10$ is more complicated, but also here a partial melting of the CDW state is realized, which continues over times longer than the ones treated by us.
Note that the parameters of the effective Hamiltonian are deeper in the CDW phase and in equilibrium one would expect a larger CDW order parameter.
In contrast, here we observe a melting of the order, which is due to the nonequilibrium protocol applied and the absorption of energy.
Driven systems in the long-time limit will realize an infinite-temperature state~\cite{lazarides_PRE2014,dalessio_PRX2014}.
In our case, as can be seen in Fig.~\ref{fig:tevol_en_orderparam}(b), the energy continues to increase as a function of time indicating that, on the transient time scale treated by us, the infinite temperature state is not yet realized.
Clearly, energy absorption is increased closer to resonance.

\begin{figure}[t!]
\includegraphics[width=0.5\textwidth]{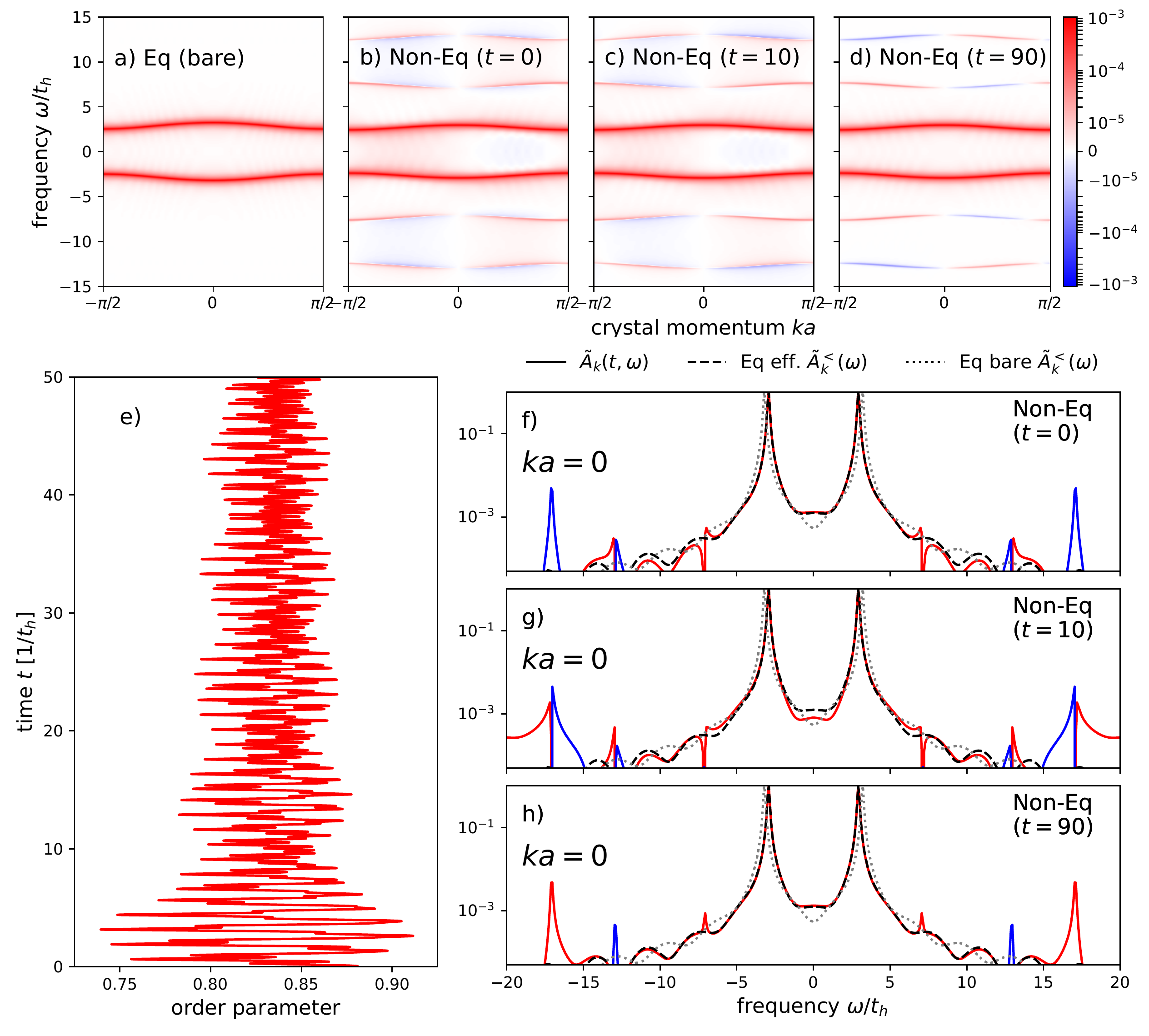}
\caption{Simulation results for the $A$-$B$ model in \eqref{eq:a_b_model} with $U = 5 t_\text{h}$ and semi-infinite sinosoidal driving with amplitude one and frequency $\Omega = 10 t_\text{h}$.
The panels a) to d) shows results for the non-equilibrium spectral functions at different times.
Panel e) shows the dynamics of the CDW order parameter under driving and panels f) to h) show cross sections through the spectral functions of b) to d).
\label{fig:a_b_model_om10}}
\end{figure}

We would like to distinguish our finding further from a known effect:
In earlier works on electron-mediated CDW melting~\cite{Shen2014,Shen2014a}, the appearance of in-gap spectral weight was reported already in a pumped non-interacting fermion model as a genuine non-equilibrium effect.
Hence, the question arises, if the new in-gap band can be obtained also in a continuously driven CDW system without interactions.
To study this, we adopt the ``$A$-$B$ model'' by Shen et al.~\cite{Shen2014} at half filling,
\begin{align}\begin{split}
\hat H = &-t_h \sum_j\left(\text{e}^{i A(t)} c_j^\dagger c_{j+1}^{\phantom{\dagger}} + \text{H.c.} \right)\\
&+ \frac{U}{2} \Big( \sum_{i \in A} c_i^\dagger c_i - \sum_{i \in B} c_i^\dagger c_i \Big) ,
\label{eq:a_b_model}
\end{split}\end{align}
and apply the same semi-infinite driving protocol used for the $t$-$V$ chain.
The CDW order in the model is due to the presence of a staggered on-site potential and leads to a spectral gap of size $\Delta \approx U$.
We use the same Trotterized time-evolution as in the original work~\cite{Shen2014} and choose a step size of $10^{-6} t_h^{-1}$.
The results of the simulations for a driving frequency of $\Omega = 10 t_\text{h}$ and a gap of $U = 5 t_\text{h}$ are shown in Fig. \ref{fig:a_b_model_om10}.
The dynamics of the order parameter in Fig.~\ref{fig:a_b_model_om10}(d) is more oscillatory than in Fig.~\ref{fig:tevol_en_orderparam}, and its envelope is decreasing in time, 
indicating CDW melting.
The momentum cuts through the spectral function, however, show that this is not connected with the formation of a peak in the spectral gap.
One should note that the spectral function does not become stationary in the model but in the in-gap region the only effect appears to be a small shift (compare Fig.~\ref{fig:a_b_model_om10}(g) and (h) ).

\begin{figure}[b]
\includegraphics[width=0.5\textwidth]{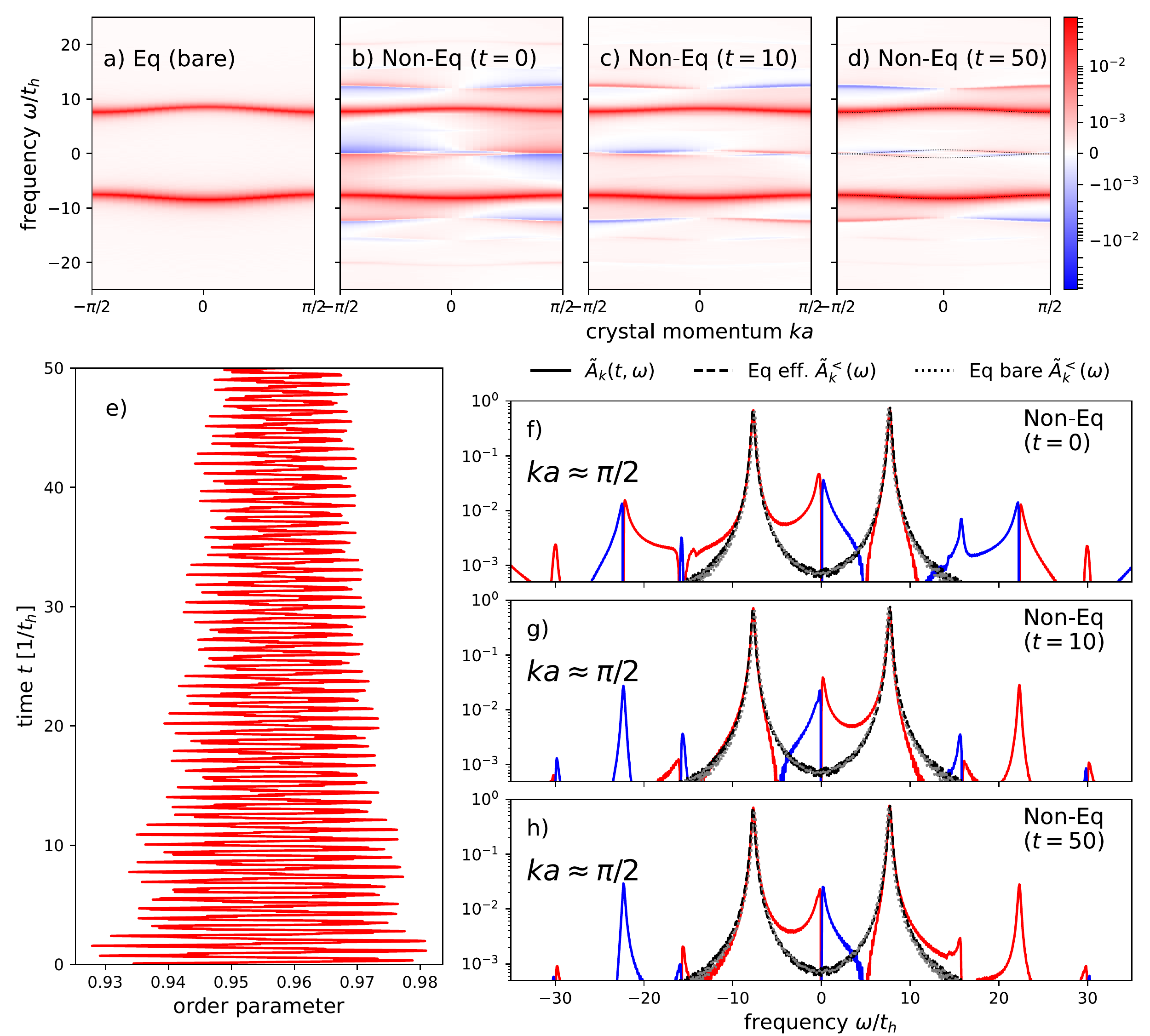}
\caption{Simulation results for the $t$-$V$ model within Hartree-Fock mean-field theory with $V = 8 t_h$  and for semi-infinite sinusoidal driving with amplitude one and frequency $\Omega = 30 t_\text{h}$.
The data is obtained for a model with $L = 64$ sites and periodic boundary conditions. The heatmaps in a) to d) are drawn for the reduced Brillouin zone.
The dotted lines in d) show the equilibrium dispersion $E_k$ and also shifted copies at $\omega = 0$ that fit well to the observed spectral weight there.
\label{fig:mf_tevol_L64_V8_om30}}
\end{figure}

To go beyond the purely non-interacting limit we treat the dynamics of the CDW phase in the $t$-$V$ model within a Hartree-Fock mean-field (MF) approach, whose results are shown in Fig.~\ref{fig:mf_tevol_L64_V8_om30}.
This will allow us to investigate the role of interaction-induced doublon excitations for the in-gap feature.
A more detailed discussion can be found in App.~\ref{app:mf_tevol}.
The equilibrium MF band structure is similar to the one in the $A$-$B$ model.
In the driven model, however, we obtain, in addition to the Floquet replicas of the equilibrium bands, a signal in the band gap around $\omega/t_h \approx 0$.
It stems from additional resonances of the band structure at separation $\Delta\omega = \pm V$, which come from the breaking of a doublon present in the system, or the formation of a doublon, respectively (a doublon here is formed by two electrons on adjacent sites). 
The larger $V$, the fewer doublons are present in the ground state, so that the resonance at $-V$ is weaker, as confirmed within our MF approach (see App.~\ref{app:mf_tevol}). 

This picture will explain in parts the MPS findings.
However, they differ from the MF results in at least two aspects: 
i) the MF results cannot reproduce the exact quasi-particle continuum and hence the in-gap resonance possesses further features caused by higher scattering processes or correlation effects.
Also, more complicated excitations can come into play, e.g., bound states, which are not captured by MF.
ii) when going closer to resonant driving, the melting of the CDW rather seems to induce a new band than a replica of existing spectral features, see Fig.~\ref{fig:specfunc_heatmap_obc64_om42}.
Taking these considerations into account, we propose the following possible scenarios leading to the in-gap band observed by MPS:

i) creation of bound states described in Ref.~\onlinecite{Pereira2009} for the system at equilibrium. 
This would lead to a $\cos$-like band whose bandwidth, however, depends on the interaction strength $\sim~1/V$.
The findings of Figs.~\ref{fig:specfunc_heatmaps_comp_eff} and~\ref{fig:specfunc_heatmaps_specfunc_heatmap_obc32_om20} for $V/t_h = 5$ and $V/t_h=10$, respectively, show that the bandwidth in our case seems to be only weakly depending on $V$, if at all. 

ii) Formation of doublons and scattering of the elementary excitations. 
A reminiscent scenario is, e.g., realized in spin-chains upon increasing the temperature \cite{spin-1,nayak_mila_PRB2022}. 
To investigate this, one should treat the single-particle excitations of the system, e.g., with Bethe ansatz, which is beyond the scope of this paper.

iii) the melting of the CDW state leads to the `emission' or creation of free carriers, which can move freely on the lattice and hence realize a tight-binding like dispersion. 
In this scenario, the incomplete melting of the CDW would lead to a situation in which remnants of the CDW crystal survive, but at the same time the system would possess also some metallic character due to the freely mobile charge carriers.

The comparison with the MF-results indicates that scenario ii) is probably best applicable deeper in the Magnus regime, while closer to resonance scenario iii) might be better suited. 
Typically, a mix will be realized. 
In addition, as indicated by Figs.~\ref{fig:panel_comp_eff_ret_t_k0},~\ref{fig:panel_comp_eff_ret_t_k25e-2} and \ref{fig:specfunc_heatmaps_specfunc_heatmap_obc32_om20} for  various parameters, there seems to be a broadening or a continuum attached to the cosine-like feature. 
Further investigations are needed to clarify this.

\section{Conclusions and Outlook}
We observe by computation of the time-dependent spectral function using MPS that the melting of an initial CDW insulator is accompanied by the formation of a new cosine-like feature in the spectral function upon periodic driving with frequencies above the band gap (``Magnus regime'').
When approaching resonance, the non-equilibrium spectral function changes significantly as compared to the equilibrium case and the new band becomes the dominant feature.
Instead, at frequencies substantially larger than the gap the original features of the spectral function are modified according to the prediction by the effective Floquet Hamiltonian~\cite{Kennes2018}, but in addition the cosine-like in-gap band prevails.
Such a feature is not observed when periodically driving a noninteracting CDW-state.
On the MF level, an in-gap band is obtained by creating or breaking a doublon, but its properties still differ significantly from the MPS results, indicating that correlation effects are needed to explain our findings. 
It will be interesting to further clarify the origin of this new band in this simple correlated system, e.g., using higher-order Floquet-Magnus expansions, Bethe ansatz, or semiclassical approaches, such as fermionic truncated Wigner approximations \cite{Davidson2017}.
It is an open question to see whether the interplay of the melting of CDW states and electron correlations can lead to similar in-gap features beyond the leading order Floquet-Magnus effective Hamiltonian also in periodically driven interacting two-dimensional systems, such as tilted bilayer heterostructures.

\acknowledgments
We thank G\"{o}tz Uhrig, Andr\'{e} Eckardt, Stefan Kehrein, Sebastian Paeckel, Thomas K\"{o}hler, Mona Kalthoff, Karlo Penc, Niklas B\"{o}lter and Karun Gadge for useful discussions. 
We are grateful for many stimulating and insightful discussions with all participants of the journal club of the B07 project of the SFB 1073, in particular also Stefan Mathias and Marcel Reutzel. 
The work was supported by the North-German Supercomputing Alliance (HLRN)
and we are grateful to the HLRN supercomputer staff.
%
%
We also acknowledge access to computational resources provided by the GWDG and acknowledge technical assistance.
%
This work is funded by the Deutsche Forschungsgemeinschaft (DFG, German Research Foundation) - 217133147/SFB 1073, projects B03 and B07.
The MPS-based results presented in this work were generated using the SymMPS toolkit~\cite{symmps}.

\appendix

\section{Additional supporting figures\label{app:supp_figures}}

\begin{figure}
\includegraphics[width=0.5\textwidth]{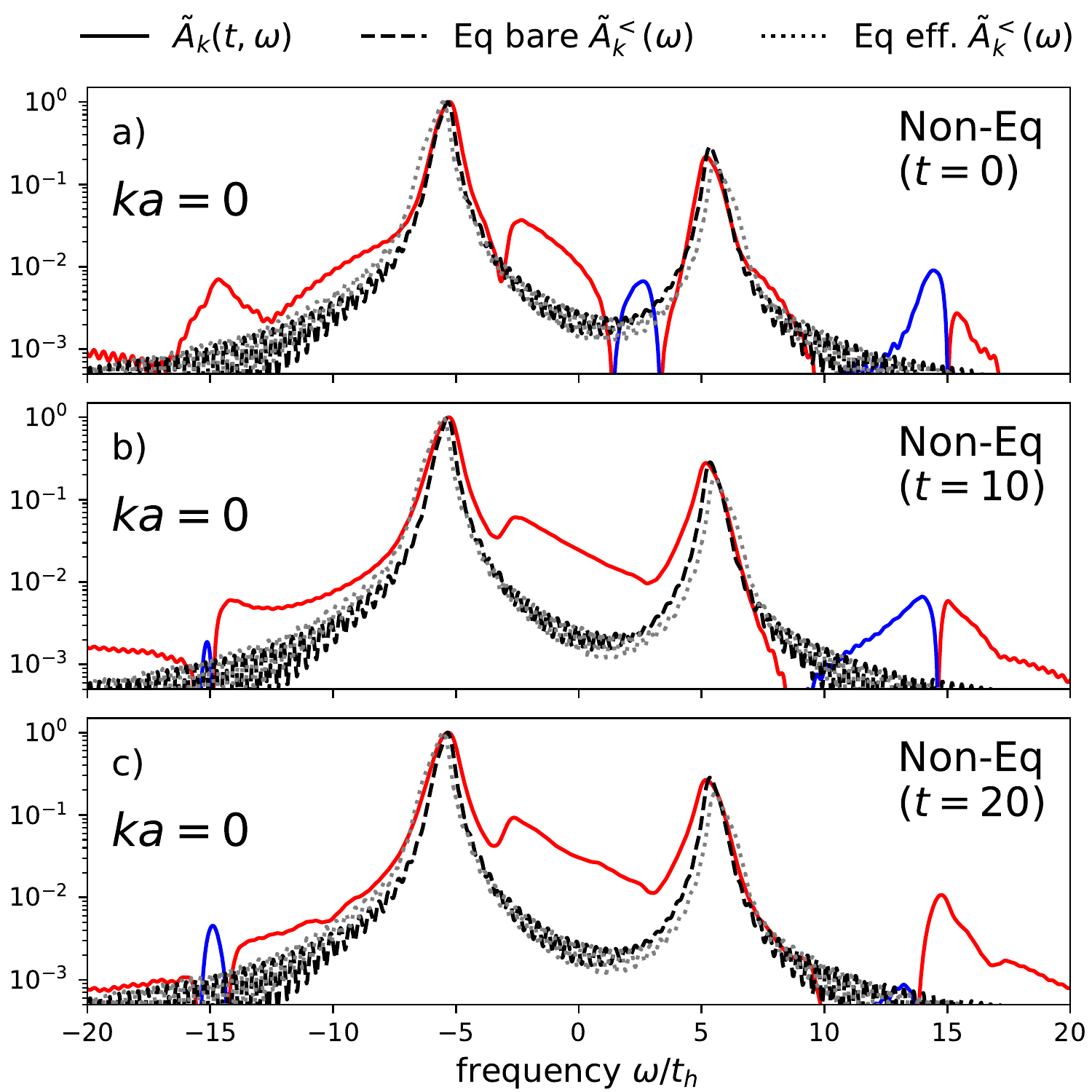}
\caption{Additional cross section through $ka \approx 0$ for the data shown in Figs. \ref{fig:specfunc_heatmaps_comp_eff} and \ref{fig:cross_sec}.
\label{fig:panel_comp_eff_ret_t_k0}}
\end{figure}

\begin{figure}
\includegraphics[width=0.5\textwidth]{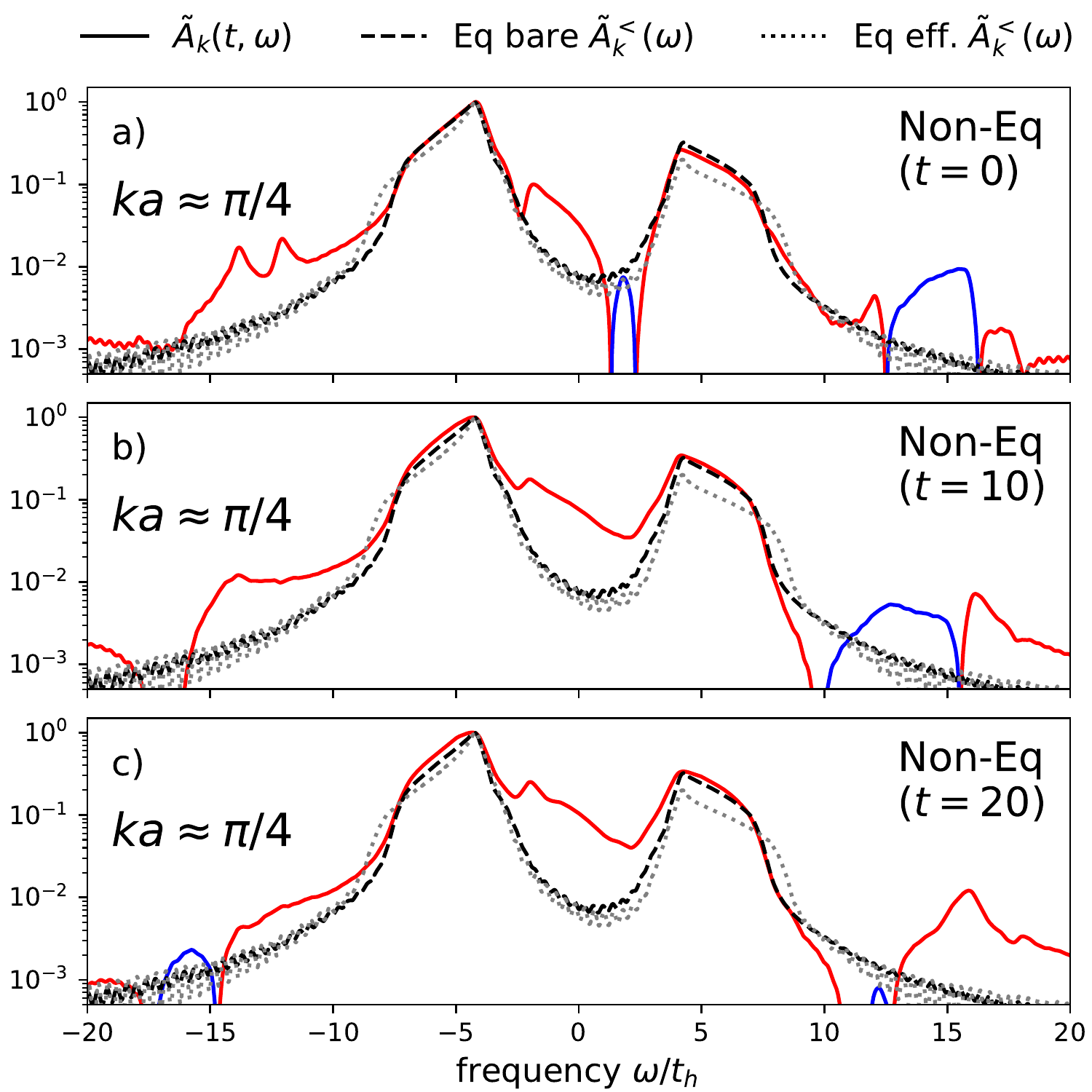}
\caption{Additional cross section through $ka \approx \frac{\pi}{4}$ for the data shown in Figs. \ref{fig:specfunc_heatmaps_comp_eff} and \ref{fig:cross_sec}.
\label{fig:panel_comp_eff_ret_t_k25e-2}}
\end{figure}

In this Appendix we present some additional figures with other parameters and more technical content.
Figs.~\ref{fig:panel_comp_eff_ret_t_k0} and \ref{fig:panel_comp_eff_ret_t_k25e-2} provide additional cross section plots at momenta $k \approx 0$ and $k \approx \pi/4$.
One can clearly identify the in-gap spectral weight as well.
It is noteworthy that here it is less sharply concentrated and a continuum of weight seems to be attached to the peak.

\begin{figure*}
\includegraphics[width=\textwidth]{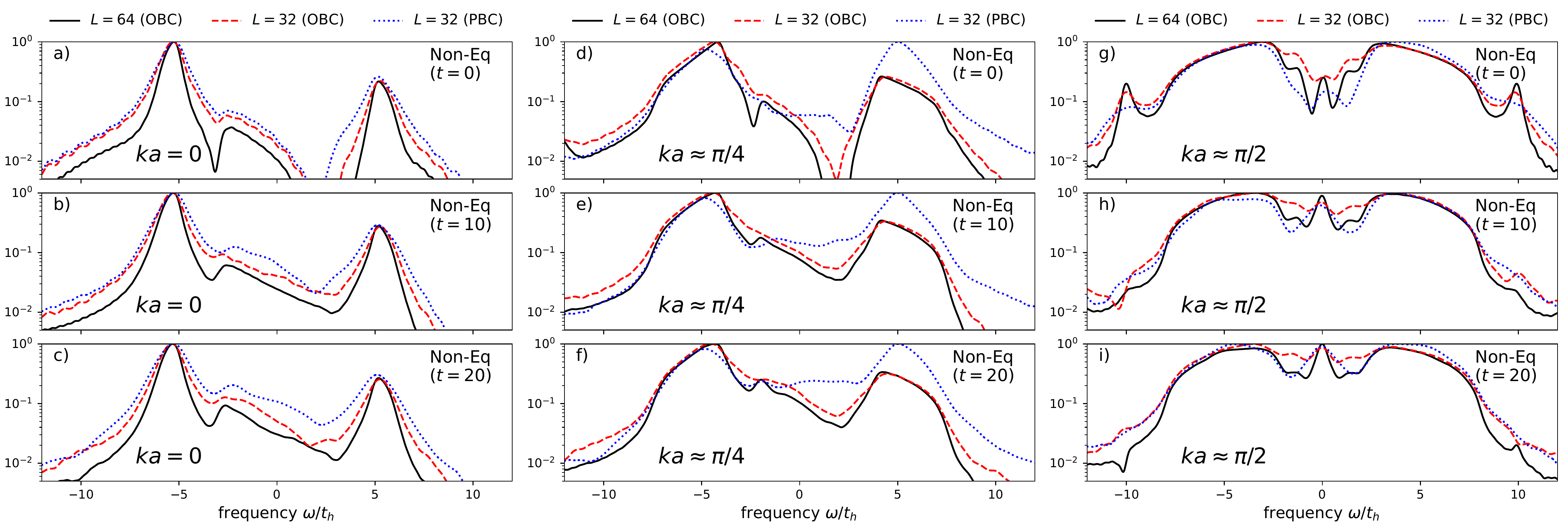}
\caption{Cross sections through the spectral functions for different chain lengths and boundary conditions.
All data sets are consistent with each other; the slice $ka \approx \pi/2$, in particular, clearly shows the additional in-gap signal.
It is more sharply resolved for the larger system ($L = 64$ sites) and for PBC in comparison with OBC and $L = 32$ sites.
As in the main text the spectral functions are normalized to their maximum value in that $k$-slice.
In these graphs only positive values of the spectral functions are shown, which covers almost all of the plotting area.
\label{fig:cross_sec_comp_Ls}}
\end{figure*}

In Fig.~\ref{fig:cross_sec_comp_Ls} we collect data for different system sizes and varying boundary conditions as a check that our observations do not depend on these parameters.
Looking at Fig.~\ref{fig:cross_sec_comp_Ls}(g) to (i) the peak at frequency zero is clearly visible for all data sets.
However, for $L = 32$ and OBC it is less pronounced.
The PBC data and OBC with 64 sites agree very well with each other.

\begin{figure*}[t]
\includegraphics[width=\textwidth]{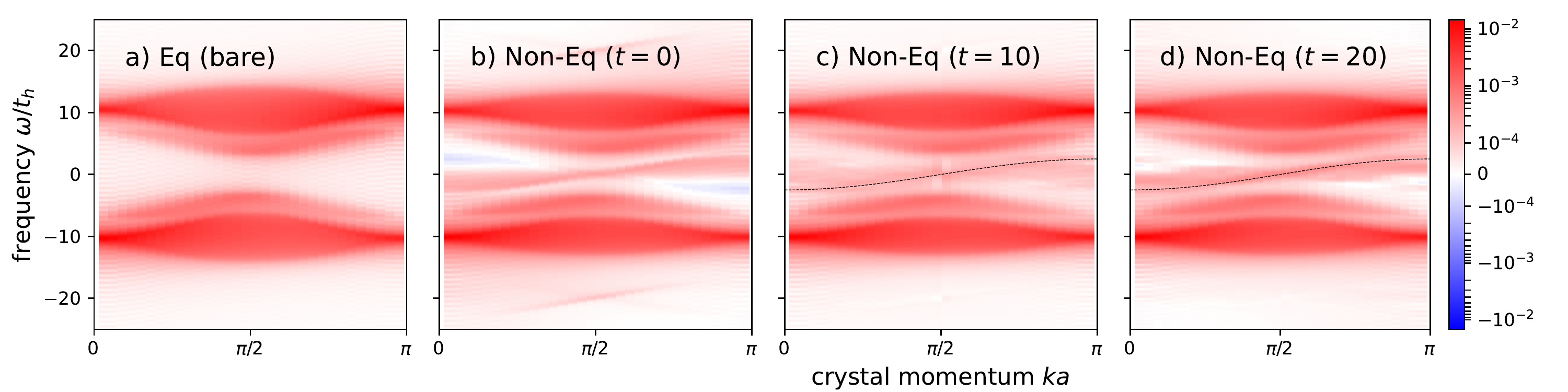}
\caption{Non-equilibrium spectral functions for double the interaction strength $V = 10 t_h$ and driving frequency $\Omega = 20 t_h$.
The spectral gap from Bethe ansatz is $\Delta/t_h \approx 6.242$.
As in the main text an in-gap spectral feature appears upon driving.
Despite the different value of $V$,
the width of the feature is similar to the one with $V = 5 t_h$.
The dotted line displays $f(k) = -2.5 \cos(ka)$ like in Fig.~\ref{fig:specfunc_heatmaps_comp_eff}
The data is obtained for OBC in a chain of $L = 32$ sites.\label{fig:specfunc_heatmaps_specfunc_heatmap_obc32_om20}}
\end{figure*}

Fig.~\ref{fig:specfunc_heatmaps_specfunc_heatmap_obc32_om20} shows a simulation with a larger value of $V = 10 t_\text{h}$ and double the driving frequency $\Omega = 20 t_\text{h}$.
Our observation from the main text is confirmed and we find the cosine-like feature again.
It follows roughly the same functional form as the signal in the main text although here the interaction strength is different (see Figure caption)

\section{Details on Green's functions, spectral functions\label{app:details_gf}}

\subsection{Definitions}

All Green's functions (GFs) are derived from the contour-ordered single-particle Green's function~\cite{stefanucci2013nonequilibrium}
\begin{equation}
 G_{\alpha \beta}(t, t') = -i \langle \hat T_C c_\alpha(t) c_\beta^\dagger(t') \rangle
\end{equation}
which can be written in a matrix representation with respect to the forward and backward branches of the real-time axis.
In this representation the greater $G_{\alpha\beta}^{>}(t, t')$ and lesser Green's function $G_{\alpha\beta}^{<}(t, t')$ each have one time argument lying on the forward and one on the backward branch of the real-time contour.
The retarded GF is a linear combination of the two with an additional theta function,
\begin{align}\begin{split}
 G_{\alpha \beta}^\text{ret}(t, t') &= \theta(t - t') \left( G_{\alpha \beta}^{>}(t, t') - G_{\alpha \beta}^{<}(t, t') \right) \\
 &:= -i \theta\left(t - t'\right) \left( \big\langle c_\alpha(t) c_\beta^\dagger(t') \big\rangle \right.\\
 &\left. \qquad\qquad\qquad + \big\langle c_\beta^\dagger(t') c_\alpha(t) \big\rangle \right) \, .
\end{split}\end{align}
At equilibrium, one of the two time variables can be suppressed due to time-translational invariance.
Out-of-equilibrium, however, we need to consider both time variables, and the Fourier transform to frequency space is not unique any more (see, e.g., Ref.~\onlinecite{Kalthoff2018}).
In order to minimize the numerical costs, we choose to use ``horizontal'' time coordinates, in which we evolve the wavefunction up to a time $t$ and then perform the Fourier transform with respect to the relative time $\tau = t - t'$ (also referred to as waiting time farther below), after further evolving the system in time, with $t > t'$
The states are labelled by momentum indices $k$ (depending on boundary conditions, see below).
In the numerics we calculate the auxiliary quantities 
\begin{align}\begin{split}
C^{<}_{k l}(t', \tau) &= \big\langle c_k^\dagger(t'+\tau) c_l(t') \big\rangle \\
C^{>}_{k l}(t', \tau) &= \big\langle c_l(t') c_k^\dagger(t'+\tau) \big\rangle \, ,
\end{split}\end{align}
such that
\begin{align}\begin{split}
G^{\text{ret},\lessgtr}_{kk}(t', \tau) = -i \theta(\tau) C^{\lessgtr}_{kk}(t', \tau)^\ast ,
\end{split}\end{align}
and we finally obtain the nonequilibrium spectral functions
\begin{align}\begin{split}
A^{\text{ret},\lessgtr}_k(t, \omega) &= -\operatorname{Im} \frac{1}{\sqrt{2\pi}} \int_{-\infty}^\infty \text{d}\tau \; e^{(i\omega - \eta) \tau} G_{kk}^{\text{ret},\lessgtr}(t, \tau) \, ,\\
A_k^{\text{ret}}(t, \omega) &= A^{\text{ret},<}_k(t, \omega) + A^{\text{ret},>}_k(t, \omega) \, .
\end{split}\end{align}

\subsection{Interpretation}

In equilibrium the retarded GF contains information about the density of states while the lesser and greater GFs contain information about occupations of the states.
The latter is reflected in the fact that due to the absence of the $\theta(\tau)$ the whole ``history'' of the lesser and greater GFs needs to be traced back until time $-\infty$.
In this text we compute the retarded Green's function.
Integration of $A_k ^{\rm ret}(t,\omega)$ over $k$ directly yields the time-dependent local density of states.
In contrast to the equilibrium sitatuation the quantity $A_k ^{\rm ret}(t,\omega)$ is not necessarily positive, so that care needs to be taken with this interpretation \cite{Kalthoff2019}.
Further issues when using the two-time Green's functions to directly model trARPES experiments arise due to the lack of gauge invariance, in particular if pump and probe pulses overlap, or if the system possesses multiple bands, see Refs.~\onlinecite{Freericks2009,Freericks2015,Freericks2016} for a detailed discussion. 
Here, we neglect these aspects and mainly focus on the most important qualitative features evolving with time.
However, central predictions of Floquet theory, like the effective Hamiltonian picture, are captured with a good accuracy, so that we believe that our results not only provide a qualitative picture, but also quantitative predictions with a good accuracy. 

\subsection{Technical aspects}

\paragraph*{Spatial Fourier transform}
For periodic boundary conditions (PBC) we use $k \in \frac{2\pi}{L} \cdot \{ 0, \dots, L \}$
while for open boundary conditions (OBC)
$k \in \frac{\pi}{L+1} \cdot \{ 1, \dots, L \}$ corresponding to a sine transform \cite{Kohler2020,Meyer2021} with $c_k = ( 2/(L+1) )^{-1} \sum_k \sin(k r_i) c_i$.
The OBC momenta differ slightly ($\sim 1/(L+1)$) from simple fractions of $\pi$, e.g. $\pi/2$, so we always consider the closest $k$-value and write~$\approx$.

\paragraph*{Temporal Fourier transform}
Due to the finite maximal $\tau$ that we are able to reach and due to the theta function,
the temporal Fourier transform produces a non-zero background $\sim 10^{-5}$ signal everywhere in the spectral function.
Since we expect the exact spectral function to have value zero if no signal is present, 
we decided to subtract this background.
Note that we add a damping factor $\eta \approx 0.1$ to regularize the finite time-propagation with respect to $\tau$.
In addition, to improve the $\omega$-resolution we padded the $\tau$-data with zeros to obtain at least 4096 frequency points.

\paragraph*{Symmetry-broken CDW state}
For MPS calculations with OBCs we study the system 
at half filling with $\mu_i = 0$.
In the CDW phase, this leads to an exact superposition of the two possible symmetry-broken ground states, so that for a finite system  $\mathcal{O}_\text{CDW}(t) = 0$ for all times $t$.
In order to be able to keep track of the dynamics of the CDW order parameter, we performed additional simulations where we applied a `pinning field' at the edge $\mu_{1,L} \neq 0$, which selects one of the ground states and allows us to study $\mathcal{O}_\text{CDW}(t)$. 
We have checked that the spectral function does not differ much if calculated with or without the pinning field.
In order to minimize possible boundary effects in the order parameter, we perform the sum in Eq.~\eqref{eq:orderparameter} only over the four unit cells in the center of the system. 

\paragraph*{MPS calculations}
In order to compute $G_{kk}^{\text{ret}, \lessgtr}(t,\tau)$ using matrix product states we start from the system's ground state $\ket{\text{GS}}$ and consider the following quantum states
\begin{align}\begin{split}
    |\psi(t)\rangle &= U_\text{TDVP}(t, 0) |\text{GS}\rangle, \\
    |\phi_{l}^{<}(t)\rangle &= c_{l} |\psi(t)\rangle, \\
    |\phi_{l}^{>}(t)\rangle &= c^{\dagger}_{l} |\psi(t)\rangle, \\
    |\phi_{l}^{\lessgtr}(t+\tau)\rangle &= U_\text{TDVP}(t + \tau, t) |\phi_{l}^{\lessgtr}(t)\rangle, 
    \label{eq:mps_states}
\end{split}\end{align}
where $U_\text{TDVP}(t, t_0)$ denotes time-evolution using the MPS realization of the time-dependent variational principle (TDVP).
Here, we always apply a two-site TDVP algorithm~\cite{Paeckel2019}.
The operators $c_k$ carry momentum space labels.
However, we always work in position space and exploit that we can write momentum space annihilation and creation operators as a sum of local operators $c_{k} = \sum_{j} P_{k,j} c_{j}$,
where we have introduced the transformation matrix $P$, allowing us to compute $C_{kl}^{\lessgtr}(t,\tau)$ through a series of local MPO-MPS applications. 
$P$ depends on the boundary conditions used.
Using the states \eqref{eq:mps_states} we calculate the quantities 
\begin{align}
    C_{kl}^{<}(t,\tau) &= \langle\psi(t+\tau)| c_{k}^{\dagger} |\phi_{l}^{<}(t+\tau)\rangle, \\
    C_{kl}^{>}(t,\tau) &= \langle\phi_{l}^{>}(t+\tau)| c_{k}^{\dagger} |\psi(t+\tau)\rangle ,
\end{align}
which are related to $G_{kk}^{\text{ret},\lessgtr}(t, \tau)$ as desctibed in the Definitions section.
For PBC we implement a ``snake geometry''~\cite{paeckel_koehler_comm} for the labelling of the sites in the chain.
It turned out that the DMRG ground state search always chose one of the degenerate ground states which allowed us to calculate the order parameter directly.
Further details on the calculation of the non-equilibrium spectral function with MPS methods can be found in Ref. \onlinecite{Meyer2021}.

\section{Hartree-Fock time-evolution\label{app:mf_tevol}}

We start from a Hartree-Fock decoupling of the interaction term and assume a two-site unit cell with sublattices $A$ and $B$.
Let us denote
\begin{align}\begin{split}
 \rho_A &:= \langle c_i^\dagger c_i \rangle_{i \in A}, \quad \rho_0 := \langle c_i^\dagger c_{i+1} \rangle_{i \in A}, \\
 \rho_B &:= \langle c_i^\dagger c_i \rangle_{i \in B}, \quad \rho_1 := \langle c_i^\dagger c_{i+1} \rangle_{i \in B} .
\end{split}\end{align}

Using the Fourier basis ($Q = \pi$)
\begin{align}\begin{split}
 c_{i \in A}^\dagger &= \frac{1}{\sqrt{V}} \sum_{k \in \text{rBZ}} \text{e}^{-i k r_i} \big( c_k^\dagger + c_{k + Q}^\dagger \big) \\
 c_{i \in B}^\dagger &= \frac{1}{\sqrt{V}} \sum_{k \in \text{rBZ}} \text{e}^{-i k r_i} \big( c_k^\dagger - c_{k + Q}^\dagger \big)
\end{split}\end{align}
we obtain, using the definitions $\epsilon_k = -2 t_h \cos(k)$, $\chi_k = V \big( \rho_0 \text{e}^{-ik} + \rho_1^\ast \text{e}^{ik} \big)$, the following representation of the Hamiltonian
\begin{widetext}
\begin{equation}
 ( H ) = \begin{pmatrix} c_k^\dagger & c_{k + Q}^\dagger \end{pmatrix} \begin{pmatrix} \epsilon_k - \operatorname{Re}(\chi_k) + V (\rho_A + \rho_B) - \mu & i \operatorname{Im}(\chi_k) + V (\rho_B - \rho_A) \\ -i \operatorname{Im}(\chi_k) + V (\rho_B - \rho_A) & -\epsilon_k + \operatorname{Re}(\chi_k) + V(\rho_A + \rho_B) - \mu \end{pmatrix} \begin{pmatrix} c_k \\ c_{k+Q} \end{pmatrix}
\end{equation}
\end{widetext}

In the following we consider half filling $\mu = V(\rho_A + \rho_B)$.
The saddle point values of $\rho_A$, $\rho_B$, etc. are determined with a simulated annealing approach.
Diagonalization of the Hamiltonian yields the eigenenergies
\begin{align}\begin{split}
 E_k &= \pm \Big\{ - \Big[ -\big( \epsilon_k - \operatorname{Re}(\chi_k) \big)^2 - \operatorname{Im}(\chi_k)^2 \\
 &\qquad\qquad - V^2 \big( \rho_B - \rho_A \big)^2 \Big] \Big\}^{1/2} \\
 &= \pm \Big\{ \big| \epsilon_k - \chi_k \big|^2 + V^2 (\rho_B - \rho_A)^2 \Big\}^{1/2} .
\end{split}\end{align}
Hence, the spectral gap is given by $2 V (\rho_B - \rho_A)$.

\begin{figure}
\includegraphics[width=0.5\textwidth]{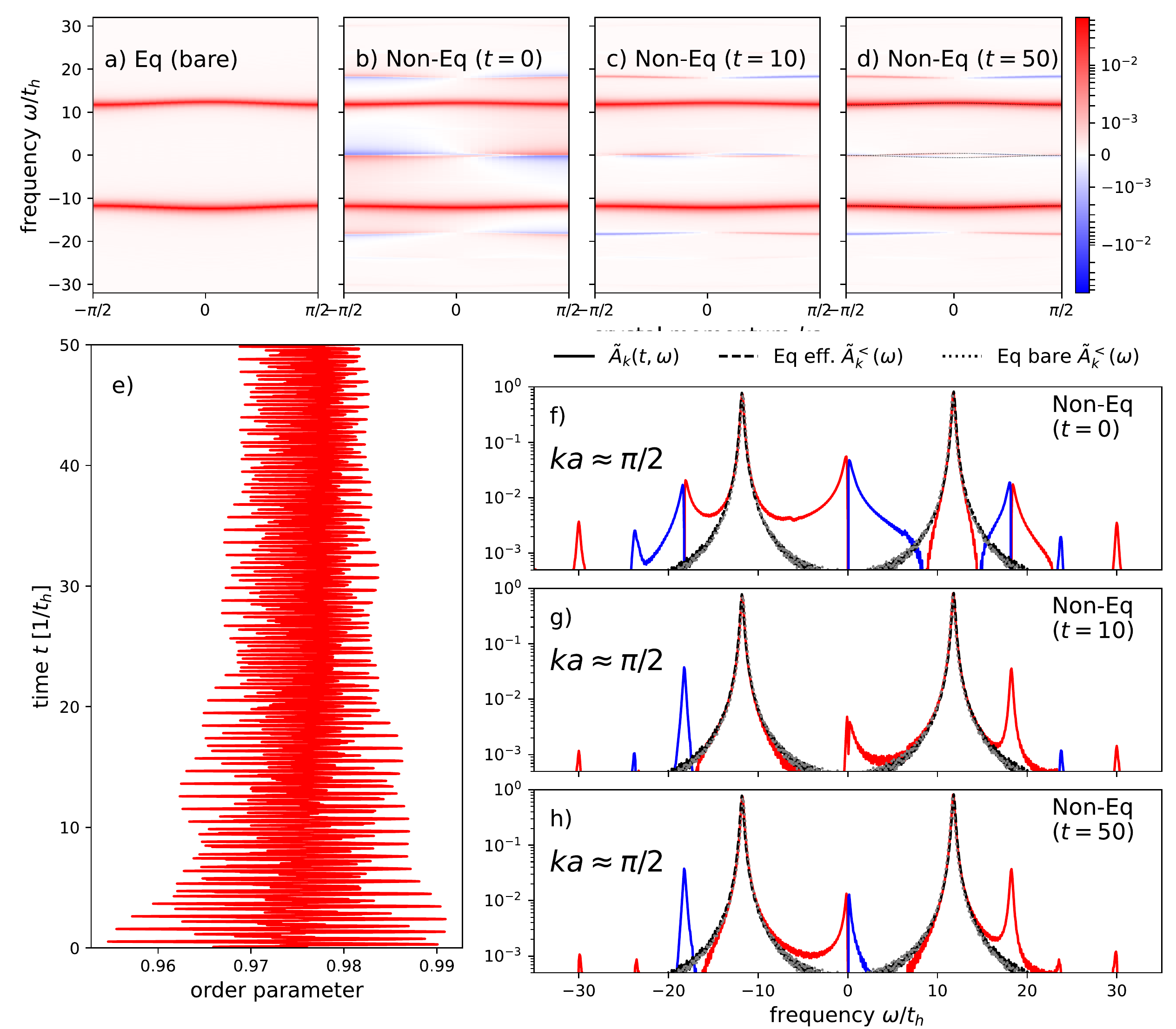}
\caption{Simulation results for the $t$-$V$ model within Hartree-Fock mean-field theory with $V = 12 t_h$  and for semi-infinite sinusoidal driving with amplitude one and frequency $\Omega = 30 t_h$ (cf. Fig.~\ref{fig:mf_tevol_L64_V8_om30})
The data is obtained for a model with $L = 64$ sites and periodic boundary conditions.
\label{fig:mf_tevol_L64_V12_om30}}
\end{figure}

For the dynamics we first solve the time-diagonal problem and obtain the full one-particle reduced density matrix $\rho_{ij}(t) = \langle c_i^\dagger(t) c_j(t) \rangle$.
In a second iteration we solve the equation of motion for the relative time $\tau$ using the time-diagonal data from the first iteration.
This corresponds to solving the Kadanoff-Baym equations with a Hartree-Fock self-energy \cite{stefanucci2013nonequilibrium}.

We have studied chains with $L = 64$ sites and periodic boundary conditions for which we calculated the time-evolution of the driven and undriven model.
For $V = 8 t_h$ (chosen to have a visual separation of $V$-dependent spectral features from Floquet sidebands) the MF spectral gap size is about $\Delta \approx 15 t_h$.
We choose a driving frequency of $\Omega = 30 t_h$, which yields the same ratio $\Delta / \Omega$ as for the $A$-$B$ model.
The results are shown in Fig.
\ref{fig:mf_tevol_L64_V8_om30}.
The undriven spectral function is very similar to the one obtained in the $A$-$B$ model although with a different spectral gap size.
The driven model, however, shows additional signals separated by $\pm V$ from the main peaks.
This gives rise to an in-gap signal around $\omega/t_h \approx 0$.
Like in the $A$-$B$ model the spectral function displays negative weights, which -- in contrast to our MPS results -- are pronounced at all times treated by us, while in the MPS case the negative weights seem to substantially decrease in time.
We consider, however, that it still contains relevant qualitative information, e.g. the position of spectral peaks.
The order parameter in Fig.~\ref{fig:mf_tevol_L64_V8_om30}(e) is oscillatory with a slowly decaying envelope.
It in fact oscillates around a larger order parameter than in equilibrium.
Still, the equilibrium CDW is broken up and charges can move in the system.
We have checked that when the driving is suddenly turned off order parameter oscillations as well as the in-gap spectral features remain.
The Floquet replicas of the main bands, however, disappear.

 \bibliography{lit}

\end{document}